\baselineskip=18pt
\overfullrule=0pt

\magnification 1200
\null
\footline={\ifnum\pageno>0   \hfil    \folio\hfil   \else\hfill   \fi}
\pageno=0

\nopagenumbers
\vskip  10pt
\centerline{\bf Guided random walk calculation of energies and $\langle \sqrt{ r^2} \rangle$ values}
\centerline{\bf   of the $^1$$\Sigma_g$ state of $H_2$ in a magnetic field.}

\vskip 10pt
\centerline{ Mario Encinosa}
\centerline {Department of Physics, Florida A \& M University}
\centerline {Tallahassee, Florida}
\vskip 4pt
\centerline{\bf Abstract}
\vskip 2pt 

Energies  and spatial observables for the $^1$$\Sigma_g$ state of the hydrogen molecule  
in magnetic fields parallel to the proton-proton axis are calculated with a
guided random walk Feynman-Kac algorithm.  We demonstrate 
that the accuracy of the results and simplicity of the method 
may prove it a viable alternative to large basis set expansions for small 
molecules in applied fields.

\vskip 2pt
\noindent
Suggested PACS numbers: 31.15 +q, 31.15 -Kb, 33.15 Bh
\vskip 4pt
\noindent
\centerline {\bf 1.Introduction}
\vskip 2pt
     The physics of the $H_2$ molecule in magnetic fields is a problem of interest in molecular
physics [1,2], astrophysics [3], and to some extent in condensed matter physics [4].
A series of papers examining $H_2$ in magnetic fields parallel to the proton-proton
axis have recently been published [5,6].  These papers use basis set expansions to compute
potential energy curves and equilibrium values of the proton-proton radial separation for several 
of the low lying states of the molecule. While this program has been  succesful,  it is 
useful to have alternate methods that have their own advantages. Work has been done on this
problem with a fixed phase Monte Carlo algorithm [7] that is general and may prove to
be the  eventual method of choice for other molecules in magnetic fields. Here however we have chosen 
a guided random walk Feynman-Kac (GRWFK) algorithm [8,9,10]  because of it's ease of 
implementation and extension to other molecular systems.  GRWFK also has the advantage 
of avoiding  trial wavefunction contamination of quantities calculated with the method.
\vskip 1pt  
     This brief report is organized as follows:  In section 2 we give an outline of the calculational 
method.  In section 3 we apply the method to the $^1$$\Sigma_g$ state of $H_2$ in a magnetic
field.  Section 4 presents results for the energies and electronic $\langle \sqrt{ r^2} \rangle$
values of the molecule for some representative fields.  Section 5 is reserved for conclusions. 
\vskip 4pt

\centerline {\bf 2. Calculational method}

We give a brief discussion of the calculational method for one dimensional systems. The extension 
to higher dimensionality is very simple and is an advantage of the formalism. 
\vskip 1pt
The Feynman-Kac formula [11,12] for a walker starting a walk at $x ^\prime, T ^\prime = 0,0$ is
$$
 U(x,T,x ^\prime = 0,T ^\prime = 0)  =  \langle exp \big [ - \int \limits_0^T ds\ V{(B(s)+x)}
 \big ]  \rangle  = \sum_n \Psi_n (x) \Psi_n (0) e^{-E_nT}.
\eqno(1)
$$
\noindent
$U(x,T,0,0)$ is the Euclidian quantum mechanical propagator and the brackets denote an expectation 
value over the Weiner measure [13].  The   argument of the exponential is an integral of the potential 
along a Gaussian process  of mean zero and variance one [14] (we choose this as a Bernoulli sequence
of plus and minus ones).  Computationally this amounts to evaluating 
$$ 
U_{one \  path}(x,T,x ^\prime = 0,T ^\prime = 0) \approx exp \lbrack -{1 \over n} \sum_{k=0}^{nT} V({B_k \over \sqrt n}
+ x) \rbrack
\eqno(2)
$$
\noindent
a large number N times. In practice the energy is better approximated by 
$$
E_0  \approx {1 \over {T_2 - T_1}} \ln {\Big[ {U(T_2)  \over {U(T_1)}} \Big ]}
\eqno(3)
$$
\noindent
with $T_1,T_2$, two sufficiently large times. 
\vskip 1pt 
It is known  the F-K formula as written in eq.(1) and it's discretized form in eq.(2) are  formally
 correct but plagued by slow convergence and large variance when employed for numerical work.  
There are many schemes in the literature that address this problem. Here we adopt a method
that has the advantage of simplicity and adaptability to parallelization. 
\vskip 1pt
An unconstrained random walk has equal probability $P = {1 \over 2}$ to step left or right. The
probability distribution $\rho (x,\tau + \Delta \tau )$ for the arrival of a walker at $x,\tau$ is 
$$
\rho(x,\tau+\Delta \tau) = {1 \over 2} \rho(x-\Delta x,\tau) +  {1 \over 2} \rho(x+\Delta x,\tau).
$$
\noindent
This leads to
$$
-{\partial \rho \over \partial \tau} = -{1 \over 2} {{h_x}^2 \over h_\tau}
{{\partial^2 \rho} \over {\partial^2 x}} .
\eqno(4)
$$
\noindent
If instead the walk probabilities for a step right or left are modified by a 
guiding function $g(x)$ such that
$$
P_{\sevenrm {L,R}} = {1 \over 2}  \pm {\partial g \over \partial x}
\eqno(5)
$$
\noindent
then in the continuum limit with  ${{h_x}^2 \over h_\tau} \rightarrow 1$ a modified diffusion equation (inclusion
of the potential $V$ is simple [8]) for the probability distribution becomes
$$
-{\partial \rho \over \partial \tau} = -{1 \over 2} 
{{\partial^2 \rho} \over {\partial^2 x}} - 2 {{\partial g }\over {\partial x}} 
 {{\partial \rho} \over {\partial x}}
+  {\rho {\partial^2 g  \over \partial^2 x}} + V \rho .
\eqno(6)
$$
\noindent
Although formally $\rho$ ought to be replaced with another symbol to account for inclusion of
$V(x)$ it should cause no confusion here.

By setting  $\rho = e^{-2g} U$ eq.(6) becomes
$$
-{\partial U \over \partial \tau} = -{1 \over 2} {{\partial^2 U} \over {\partial^2 x}}
 + \Omega U,
\eqno(7)
$$
$$
\Omega (x) = V(x)  - 2 {\Big[ {\partial g(x) \over \partial x} \Big ]}^2 +  
{\partial^2 g(x) \over \partial^2x}.
\eqno(8)
$$
\noindent
Since $g = -{1 \over 2}ln U$ (from $\rho \propto {\Psi}^2)$ knowledge of the solution
 of eq.(6) give would the exact guiding function $g(x)$ and vice-versa. 
Of course $U$ is not generally known, but by choosing $g(x)$ to incorporate what 
is known about the general character of the solution
the variance of calculated observables can be reduced
and convergence to the final result can be substantially improved.
In what follows we evaluate eq.(7) with conditional probabilities given by eq.(5), and
use $\Omega(x)$ rather than $V(x)$ as the argument of the exponential in eq.(2).  For the ground
state $U \propto \Psi_0$ so we refer to trial wavefunctions rather than propagators below.
\vskip 1pt
The ground state expectation value of an operator $O({\bf r}_1)$ can be found by evaluating
$$
\langle O \rangle =  {{{\sum_{{\bf r}_1} O({\bf r}_1) {\vert \Psi({\bf r}_1) \vert}^2}} \over
                   {{\sum_{{\bf r}_1} {\vert \Psi({\bf r}_1) \vert}^2} }}.
$$
For large times $T_1,T_2$ the rightmost term of eq. (1) can be used to generate 
${\vert \Psi ({\bf r}_1 ) \vert}^2$, hence $\langle O \rangle$:  Evolve the walk to 
to time $T_1$, tallying the value of the exponential of eq.(2) for each path   at $T_1$. 
Multipy this value by $\langle O(T_1) \rangle$,then evolve the walk to $T_2$.  With
$$
w_i =  exp \big [ - \int \limits_0^{T_2} d{\bf s} \space \Omega({\bf s}) \big ]_i
$$ 
\vskip 1pt
\noindent
this yields
$$
\langle O \rangle = { \sum_{i=1}^N w_i {\big [{O(T_1) \big ]}_i} \over {\sum_{i=1}^N w_i}}.
$$
It is worth noting that we do not generate random walks on a three dimensional
grid for each particle but rather as three separate one dimensional grids.
This is why the three dimensional spatial branching factor
of  6  used by Barnes et.al. does not appear above.
\vskip 4pt

\centerline {\bf 3. Application to $H_2$ in a magnetic field.} 
\vskip 2pt
In the clamped nuclei approximation the Hamiltonian of the $H_2$ molecule in a 
magnetic field $B$  may be written in natural units as
$$
 \sum_{i=1}^{2}\bigg [ -{1 \over 2}{\nabla^2}_i  
 - {1 \over {\big | {{\bf r}_i - {{\bf R} \over 2}}} \big |}
 - {1 \over {\big | {{\bf r}_i + {{\bf R} \over 2}}} \big |}
+ {1 \over 8} \big [{{{\bf B} \over B_0} \times {\bf r}_i} \big ]^2 \bigg ]
+ {1 \over 2}{{{\bf B} \over B_0} \cdot {\bf L}} 
+ {1 \over {\big |{{\bf r}_1 - {\bf r}_2}} \big|}
+ {\bf S} \cdot {{\bf B} \over B_0}
+ {1 \over R} 
\eqno(9)
$$
\vskip 1pt
\noindent
with $B_0 = 2.3505 \times 10^9 G$. Taking $B$ along the z axis 
reduces eq. (9) for the  $^1$$\Sigma_g$ state to
$$
 \sum_{i=1}^{2}\bigg [ -{1 \over 2}{\nabla^2}_i  
 - {1 \over {\big | {{\bf r}_i - {{\bf R} \over 2}}} \big |}
 - {1 \over {\big | {{\bf r}_i + {{\bf R} \over 2}}} \big |}
+ {1 \over 8} {\gamma}^2 \big ( {x_i}^2 + {y_i}^2 \big ) \bigg ]
+ {1 \over {\big |{{\bf r}_1 - {\bf r}_2}} \big|}
+ {1 \over R},
\eqno (10)
$$
$$
\gamma = {B \over B_0}.
$$
\vskip 1pt
\noindent
For the trial  $^1$$\Sigma_g$ wavefunction it is convenient to first define the 
auxilary quantities
$$
{f_i}^{\pm} =  -{\big |}{\bf r}_i \pm {{\bf R} \over 2} {\big |}.
$$
\vskip 1pt
\noindent
The trial wavefunction  is then
$$
\Psi_t({\bf r}_1,{\bf r}_2,{\bf R}) =
\bigg [ e^{{f_1}^+}+e^{{f_1}^-} \bigg ] 
\bigg [ e^{{-{\gamma \over 4}} ({x_1}^2 + {y_1}^2)} \bigg ]
 \times   \bigg [1 \rightarrow 2 \bigg ]
\eqno(11)
$$
\vskip 1pt
It is straightforward to insert the $\Psi_t$ of eq. (11)  into the 
expression for $g$ to obtain  $\Omega$ and the modified walk probabilities. 
We note in passing that this choice amounts to a baseline level trial
function.  No a priori attempt was made to include free parameters for later
optimization.

\vskip 4pt
\centerline {\bf 4. Results}
\vskip 2pt
In table 1 we give $E_0$ in atomic units for several values of $\gamma$ at the 
indicated $R_{eq}$.  We also give  $E_0$ for $\gamma = 2.127207$ at $R_{eq} = 1.07$ a.u. 
There are two values in the literature for $E_0$ at this field strength [6,15].
Our value  confirms the result of Detmer et. al. in [6]. 
\vskip 1pt
 These results show good agreement with [6] for the relatively small number of sample paths considered
here. The value  n = 800 was arrived at  by starting at n = 200 and increasing n by 100 steps per run for 
a few thousand sample paths.   n = 500 was found to be sufficient for all $\gamma$ considered.  
300 more steps were added as a check on the stability of convergence. We give the results for
the larger n.  A similar method was employed for the determination of $T_1$ and $T_2$. 
(We did not include disassociation values since they were also the same as those given in [6].)
In table 2 we show results for calculations of $\langle \sqrt{ r^2} \rangle$ for some values of $\gamma$. We were 
interested in the value of the field for which the rms value equaled the proton's separation
believing this to be a reasonable albeit naive  measure of the onset of the transition to 
the free atomic limit.   Clearly the field value  at which this would occur is large enough
to align the spins making the triplet rather than the singlet state the correct 
state to investigate. It
should be noted that the times $T_1$ and $T_2$ that give  convergence of $E_0$ usually were
 not sufficient to converge $\langle \sqrt{ r^2} \rangle$.  This is reflected in the larger
values of $T_1$ and $T_2$ in table 2.
\vskip 4pt
\centerline {\bf 5. Conclusions}

The  accuracy of the GRWFK method for this problem suggests that faster convergence with
basis set expansions might be achieved with LCAO trial wavefunctions functions multiplied by oscillator
eigenstates. This is a direction best explored by those working already with that method.  Certainly 
basis sets expansions prove (at least for fixed nuclei) a superior method for digit aquisition
than Monte Carlo methods.  The trade off between accuracy and ease of implementaion for molecules
like  $H_3$ or even $H_3$$^+$  in a magnetic field is an open question.  
\vskip 1pt
Calculating the properties of low-lying excited states is not unfeasable but is known to be
limited by the (lack of) knowledge of the nodal structure of the state [16,17].  The 
fixed node method [18,19] has been shown as a good approximation for small molecular
systems [20].  For biased random walks the nodal structure can be built into the the 
trial functions although not without increased  complexity. 
Clearly  states  of non-zero orbital angular momentum will have to be considered
by incorporating  the techniques discussed in [7] to GRWFK. This is a topic of 
current investigation.
\vfill \eject

\vskip 6pt
\centerline{\bf Acknowledgments}
\vskip 4pt
The author would like to acknowledge the Center
 for Nonlinear and Nonequilibrium Aeroscience (CeNNAs)  for
partial financial and computing support.

\vfill\eject

\vskip 20pt
\centerline{\bf REFERENCES}
\vskip 6pt
\noindent
1. A.V. Korolev and M.A. Liberman, Phys. Rev. Lett. ${\bf 74}$, 4096 (1995).
\vskip 6pt
\noindent 
2. A.V. Korolev and M.A. Liberman, Phys. Rev. A ${\bf 45}$, 1762 (1992).
\vskip 6pt
\noindent
3. M.C. Miller and D. Neuhauser, Mon. Not. R. Astron. Soc. ${\bf 253}$, 107 (1991).
\vskip 6pt
\noindent
4. V.B. Timofeev and A.V. Chernenko,JETP Lett.  ${\bf 61}$, 617 (1995).
\vskip 6pt 
\noindent
5.  T.Detmer, P. Schmelcher, F.K. Diakonos, and L.S. Cederbaum, 
Phys. Rev. A ${\bf 56}$, 1825 (1997).
\vskip 6pt
\noindent
6. Yu. P. Kravchenko and M.A. Liberman, Phys. Rev. A ${\bf 57}$, 3404 (1998).  
\vskip 6pt
\noindent
7. G. Ortiz, M.D. Jones, and D.M. Ceperley, Phys. Rev. A ${\bf 52}$, R3405 (1995).
\vskip 6pt
\noindent
8. S.A. Chin, J.W. Negele, and S.E. Koonin, Ann. of Phys. ${\bf 157}$ 
\vskip 6pt
\noindent
9. T. Barnes and G.J. Daniell, Nucl. Phys. B ${\bf 257}$, 173 (1985).
\vskip 6pt
\noindent
10. T. Barnes and D Kotchan , Physical Review D ${\bf 35}$,1947 (1987). 
\vskip 6pt
\noindent
11. M. Kac, Trans. Am. Math. Soc., ${\bf 65}$, 1 (1949).
\vskip 6pt
\noindent
12.G. Roepstorff , ${\it Path \ Integral \ Approach \ to
\ Quantum \ Mechanics}$, (Springer Verlag, Berlin 1994).
\vskip 6pt
\noindent
13. P.K. MacKeown, ${\it Stochastic \ Simulation \ in  \ Physics}$,
(Springer Verlag, Singapore Pte. Ltd. 1997).
\vskip 6pt 
\noindent
14. R. Iranpour and P. Chacon, ${\it Basic \ Stochastic \ Processes}$,
(McMillan Publishing Co., New York, 1988).
\vskip 6pt
\noindent
15.A. V. Turbiner,JETP Lett.  ${\bf 38}$, 618 (1983).
\vskip 6pt
\noindent
16. D.J. Klein and H.M. Pickett, J. Chem. Phys. ${\bf 64}$, 4811 (1976).
\vskip 6pt
\noindent
17. D. Ceperley, J. Stat. Phys., ${\bf 63}$, 1237 (1991).
\vskip 6pt
\noindent
18. D. Ceperley and M.H. Kalos, in ${\it Monte \ Carlo \ Methods \ in \ Statistical
\ Physics}$, K. Binder, Ed. (Springer-Verlag Berlin, 1979).
\vskip 6pt
\noindent
19.  M. Caffarel and P. Claverie, J. Chem. Phys., ${\bf 88}$, 1100 (1988).
\vskip 6pt
\noindent
20. P.J. Reynolds, D. Ceperley, B. Adler and W. Lester, J. Chem. Phys. ${\bf 77}$, 5593 (1982).

\vfill\eject

\vskip 12pt
\hsize=3.5in
\noindent
{\bf Table 1} $\sp 1$$\Sigma_g$ state energies as functions of magnetic field $B$ and
$R_{eq}$.  All units are in a.u.  Bracketed numbers are last digit uncertainty
estimates  from  variance calculations for $N=50000$ sample paths. $n = 800,T_1=7,T_2=8$ for 
all entries.  
\vskip 10pt
\hrule
\vskip 2pt
\hrule
\vskip 6pt
\settabs 4 \columns
\+&&Ref. [6] &This work\cr
\vskip 6pt
\hrule
\vskip 3pt
\+\ \ $\gamma$&\ $R_{eq}$&\ \ \ $E_0$& \ \ \ $E_0$\cr
\vskip 3pt
\hrule
\vskip 6pt
\+.01&1.40&-1.173436&-1.172(2)\cr
\+.10&1.39&-1.169652&-1.169(1)\cr
\+1.0&1.24&-0.890336&-0.889(1)\cr
\+10.0&.70&\ \ 5.88902&\ \ 5.888(3)\cr
\vskip 6pt
\hrule
\vskip 2pt
\hrule

\vskip 60pt
\hsize=3.5in
\noindent
{\bf Table 2} $\sp 1$$\Sigma_g$ $\sqrt{r^2}$ values in a.u.for some values of $B$.
Data below are for  $n = 800,T_1=10,T_2=12$ and  $N=50000$ sample paths. Larger
times are needed to converge $\sqrt{r^2}$ compared to those needed for $E_0$. The 
last entry in the table serves as a check on the dissacociative limit.
  
\vskip 10pt
\hrule
\vskip 2pt
\hrule
\vskip 6pt
\settabs 3 \columns
\vskip 3pt
\+\ \ $\gamma$&\ $R_{eq}$& $\sqrt{r^2}$\cr
\vskip 3pt
\hrule
\vskip 6pt
\+.01&1.40&1.565(3)\cr
\+1.0&1.24&1.326(2)\cr
\+2.0&1.09&1.141(3)\cr
\+5.0&0.86&0.882(2)\cr
\+7.5&0.74&0.782(1)\cr
\+10.0&0.70& 0.658(2)\cr
\vskip 6pt
\hrule
\vskip 6pt
\+0.0&50.0&1.495(2)\cr
\vskip 6pt
\hrule
\vskip 2pt
\hrule

\end